\documentclass[aps,showpacs,twocolumn,amsmath,amssymb,floatfix]{revtex4}
\usepackage{graphicx,eucal}
\begin{document}
\title{Statistics of Partial Minima}
\author{E.~Ben-Naim$^1$}
\email{ebn@lanl.gov}
\author{M.~B.~Hastings$^1$}
\email{hastings@lanl.gov}
\author{D.~Izraelevitz$^2$}
\email{izraelevitz@lanl.gov}
\affiliation{
{}$^1$Theoretical Division and Center for Nonlinear Studies,
Los Alamos National Laboratory, Los Alamos, New Mexico 87545 USA\\
{}$^2$Decision Applications Division,
Los Alamos National Laboratory, Los Alamos, New Mexico 87545 USA}
\begin{abstract}
Motivated by multi-objective optimization, we study extrema of a set
of $N$ points independently distributed inside the $d$-dimensional
hypercube.  A point in this set is $k$-dominated by another point when
at least $k$ of its coordinates are larger, and is a $k$-minimum if it
is not $k$-dominated by any other point. We obtain statistical
properties of these partial minima using exact probabilistic methods
and heuristic scaling techniques.  The average number of partial
minima, $A$, decays algebraically with the total number of points,
\hbox{$A\sim N^{-(d-k)/k}$}, when $1\leq k<d$.  Interestingly, there
are $k-1$ distinct scaling laws characterizing the largest coordinates
as the distribution $P(y_j)$ of the $j$th largest coordinate, $y_j$,
decays algebraically, \hbox{$P(y_j)\sim (y_j)^{-\alpha_j-1}$}, with
\hbox{$\alpha_j=j\frac{d-k}{k-j}$} for $1\leq j\leq k-1$.  The average
number of partial minima grows logarithmically, \hbox{$A\simeq
\frac{1}{(d-1)!}(\ln N)^{d-1}$}, when $k=d$. The full distribution of
the number of minima is obtained in closed form in two-dimensions.
\end{abstract}

\maketitle

A host of decisions in computer science, economics, politics, and
everyday life involve multiple criteria or multiple objectives
\cite{rs,snt,ft,or}. A pedestrian choosing a walking path considers
the distance, the number of turns, and the number of traffic lights.
In business, takeover bids are decided on a multitude of complex
conditions in addition to the total monetary offer.  In elections,
voters examine how candidates stand on multiple issues.

In multi-objective optimization, a solution that is optimal with
respect to all criteria is rarely possible and instead, one faces a
set of choices that are suboptimal on most criteria.  Decisions
require algorithms to weed out inferior choices, sort through all the
remaining imperfect choices, and evaluate their overall quality.

Motivated by multi-criteria decision problems, we study the statistics
of multi-variate imperfect minima.  We consider a set of $N$ points in
$d$-dimensions, with coordinates ${\bf x}\equiv (x_1,x_2,\ldots,x_d)$.
Each coordinate $x_i>0$ is a distinct cost and by convention,
small-$x$ values are superior and are considered {\it dominant}.

Partial minima, analog to imperfect choices, are defined as follows.
A point ${\bf x}$ is said to be {\it $k$-dominated} by ${\bf x}'$ when
at least $k$ of the coordinates of ${\bf x}$ are larger then the
corresponding coordinates of ${\bf x}'$.  A point is said to be a
partial minimum, or formally a {\it $k$-minimum}, when it is not
$k$-dominated by any other point in the set.  We stress that a partial
minimum is not required to dominate all other points on the same $d-k$
coordinates and may dominate different points along different
coordinates.  The parameter $1\leq k\leq d$ quantifies the quality of
the partial minimum: a smaller $k$ value represents a more stringent
condition.  The two extremes are the perfect minimum, $k=1$, where
every coordinate is a minimum of the point set, and the efficient set,
$k=d$, that includes all points that are not obviously dominated by
other points as shown later in figure 1. Partial minima are
conditional multivariate extrema and their properties are amenable to
analysis using a statistical physics perspective
\cite{sg,dl,bkm,ktnr,kr}.

In this study, we obtain exact statistical properties of partial
minima including the multivariate density and its asymptotic behavior
as well as scaling properties such as the typical size and average
number of minima.  We present two major results. First, as a function
of the set size $N$, the average number of minima decays algebraically
when $1\leq k<d$, and grows logarithmically when $k=d$.  Second, there
are $k-1$ different scaling laws for the largest coordinates, each
following a power-law distribution with $k-1$ distinct exponents. The
rest of the $d+1-k$ coordinates are characterized by distributions
with sharp tails. We also discuss the relevance of these results to
the multi-objective shortest path on graphs, a central problem in
multi-objective optimization.

We consider the situation where there are no correlations between the
coordinates. That is, each coordinate is independently drawn from some
distribution. As discussed below, this situation is equivalent to a
uniform distribution in the unit hypercube. Thus, we conveniently
assume that $x_i$ is uniformly distributed in $[0:1]$ for all $1\leq i
\leq d$.

\noindent{\bf Heuristic Arguments.} Elementary scaling laws for the
typical size of a partial minimum and the average number of minima are
derived heuristically. We assume that (i) the partial minimum is
dominant on a fixed set of $k$ coordinates, and (ii) all its
coordinates are equal, $x_i=x$, for all $i$. By the partial minimum
definition, the corresponding $k$-dimensional hypercube contains only
the partial minimum itself.  The volume of this hypercube is $x^k$ and
the expected number of points inside this hypercube must be of order
one, $Nx^k\sim 1$. Consequently, the typical size $x$ decays
algebraically with $N$,
\begin{equation}
\label{scale}
x\sim N^{-\frac{1}{k}}.
\end{equation}
This characteristic scale decreases as the minimum condition becomes
more stringent, that is, as $k$ decreases.

The expected number of partial minima
\begin{equation}
\label{average}
A\sim N^{-\frac{d-k}{k}}
\end{equation}
follows from the expected number of points inside the $d$-dimensional
hypercube with linear dimension $x$, $Nx^d$. Partial minima are
asymptotically rare and the scale \eqref{scale} decays
indefinitely. Furthermore, with a small probability, there is only one
minimum when $N$ is large. The scaling estimate \eqref{average}
coincides with the exact value $A=N^{-(d-1)}$ for $k=1$, since any
point is a perfect minimum with probability $N^{-d}$. For $k=d$, the
minimum in any one coordinate is a partial minimum and thus, there is
at least one partial minimum.  Indeed, the decay exponent
$\frac{d-k}{k}$ in \eqref{average} vanishes. This special case is
discussed separately.

\noindent{\bf The Density of Minima.}  The density $P_{d,k}({\bf x})$
of $k$-minima located at ${\bf x}$ is obtained analytically through a
formal generalization of the heuristic argument above. For example, in
two dimensions the density is
\begin{eqnarray*}
P_{2,k}(x_1,x_2)=
\begin{cases}
N\left[1-\left(x_1+x_2-x_1x_2\right)\right]^{N-1}&k=1,\\
N\left[1-x_1x_2\right]^{N-1}&k=2.
\end{cases}
\end{eqnarray*}
The factor $N$ is the number of ways to choose the minimum, and the
second factor guarantees that the rest of the points do not dominate
the minimum at $(x_1,x_2)$. These points must not fall inside an
$L$-shaped region of area $x_1+x_2-x_1x_2$ or equivalently
$1-(1-x_1)(1-x_2)$ when $k=1$ or a rectangle of area $x_1x_2$ when
$k=2$.

In general, the density of minima
\begin{eqnarray}
\label{density}
P_{d,k}({\bf x})=N[1-G_{d,k}({\bf x})]^{N-1}.
\end{eqnarray}
reflects that the $N-1$ points are excluded from a $d$-dimensional
region of volume $G_{d,k}({\bf x})$. The excluded volume obeys the
recursion
\begin{equation}
\label{g-rec}
G_{d,k}({\bf x})=x_d\,G_{d-1,k-1}({\bf x})+(1-x_d)\,G_{d-1,k}({\bf x}).
\end{equation}
In our notation, the dimensional index of a function dictates the
dimension of its vectorial argument so the vectors on the right hand
side of \eqref{g-rec} have $d-1$ components. We obtain the recursion
relation \eqref{g-rec} by separating the excluded region into two
regions: one in which the $d$th coordinate is dominant and one in
which it is not. Using the boundary conditions $G_{d,0}=1$ and
$G_{d,k}=0$ when $k>d$, we recover $G_{1,1}=x_1$ and
\hbox{$G_{2,1}=x_1+x_2-x_1x_2$}. Furthermore,
\begin{eqnarray*}
G_{3,k}\!=\!
\begin{cases}
x_1\!+\!x_2\!+\!x_2\!-\!x_1x_2\!-\!x_1x_3\!-\!x_2x_3\!+\!x_1x_2x_3&k=1,\\
x_1x_2\!+\!x_1x_3\!+\!x_2x_3\!-\!2x_1x_2x_3&k=2,\\
x_1x_2x_3&k=3.\\
\end{cases}
\end{eqnarray*}
In general, $G_{d,d}=\prod_{i=1}^d x_i$ and $G_{d,1}=1-\prod_{i=1}^d
(1-x_i)$.

\noindent{\bf Scaling.} In the limit $N\to\infty$, the product term
$x_1x_2$ in \hbox{$P_{2,1}=N[1-(x_1+x_2-x_1x_2)]^{N-1}$} is negligible
compared with the linear term $x_1+x_2$ and thus,
\begin{eqnarray*}
P_{2,1}(x_1,x_2)\to Ne^{-N(x_1+x_2)}.
\end{eqnarray*}
Generally, only the $k$th degree terms are asymptotically relevant and
the leading behavior is
\begin{equation}
\label{leading}
P_{d,k}({\bf x})\to Ne^{-NF_{d,k}({\bf x})}.
\end{equation}
The auxiliary function $F_{d,k}({\bf x})$ contains ${d\choose k}$
terms, each a distinct product of degree $k$. For example,
\begin{eqnarray*}
F_{3,k}\!=\!
\begin{cases}
x_1+x_2+x_3&k=1,\\
x_1x_2+x_1x_3+x_2x_3&k=2,\\
x_1x_2x_3&k=3.\\
\end{cases}
\end{eqnarray*}
The auxiliary function equals the sum, \hbox{$F_{d,1}=\sum_{i=1}^d
x_i$}, and the product, \hbox{$F_{d,d}=\prod_{i=1}^d x_i$}, in the two
extremes. The function $F_{d,k}({\bf x})$ is defined recursively
\begin{equation}
\label{f-rec}
F_{d,k}({\bf x})=x_dF_{d-1,k-1}({\bf x})+F_{d-1,k}({\bf x})
\end{equation}
for $1\leq k\leq d$ with the boundary condition
$F_{0,k}=\delta_{k,0}$. This recursion follows from \eqref{g-rec} by
dropping the higher-degree term $x_d\,G_{d-1,k}({\bf x})$.

The asymptotic behavior \eqref{leading} can be recast in the scaling
form
\begin{equation}
\label{scaling-form}
P_{d,k}({\bf x})\to N\Phi_{d,k}({\bf z}),
\end{equation}
as $N\to\infty$. The scaling variable is ${\bf z}={\bf x}N^{1/k}$, in
accord with \eqref{scale}, and the scaling function is
\begin{equation}
\label{scaling-function}
\Phi_{d,k}({\bf z})=e^{-F_{d,k}({\bf z})}.
\end{equation}

The average number of $k$-minima equals the integral of the density,
$A_{d,k}=\int d{\bf x}\, P_{d,k}({\bf x})$, where \hbox{$\int d{\bf
x}\equiv \prod_{i=1}^d \int_0^1 dx_i$} \cite{rec}.  When $k<d$, the
asymptotic behavior of the average follows from the scaling form
\eqref{scaling-form},
\begin{equation}
\label{tot}
A_{d,k}\simeq a_{d,k}\,N^{-\frac{d-k}{k}},
\end{equation}
and is in agreement with \eqref{average}. The proportionality constant
$a_{d,k}$ equals the integral of the scaling function,
\hbox{$a_{d,k}=\int d{\bf z}\, \Phi_{d,k}({\bf z})$}, although now,
the integration range is unrestricted, \hbox{$\int d{\bf z}\equiv
\prod_{i=1}^d \int_0^\infty dz_i$}. The prefactor is trivial for
perfect minima, $a_{d,1}=1$, and otherwise, it can be obtained
analytically only in a few exceptional cases including for example
\hbox{$a_{3,2}=\frac{1}{2}\pi^{3/2}$}.

\noindent{\bf Extreme Statistics.} By definition, partial minima may
be dominant on certain coordinates but inferior on others. We
therefore study extremal statistics \cite{ejg,jg,res} to investigate
possible disparities between the coordinates.

First, consider the largest coordinate.  Without loss of generality,
we order the coordinates \hbox{$x_1<x_2<\cdots<x_{d-1}<x_d$}.  Our
focus is on the tail of the distribution of the variable $x_d$,
corresponding to the regime $x_d\gg x_{d-1}$. We also restrict our
attention to the limit $N\to\infty$.  The distribution $Q_1(x_d)$ of
the largest coordinate $x_d$ equals the integral of the multivariate
distribution with respect to the rest of the coordinates,
\begin{eqnarray}
\label{largest}
Q_1(x_d)
&=&\int dx_1\cdots \int dx_{d-1}\, P_{d,k}(x_1,x_2,\cdots,x_d)\nonumber\\
&\sim&\int dx_1\cdots \int dx_{d-1}\, Ne^{-NF_{d,k}({\bf x})}\nonumber\\
&\sim&\int dx_1\cdots \int dx_{d-1}\, Ne^{-Nx_dF_{d-1,k-1}({\bf x})}\nonumber\\
            &\sim&N^{-\frac{d-k}{k-1}}(x_d)^{-\frac{d-k}{k-1}-1}.
\end{eqnarray}
The second line is obtained by substituting the leading asymptotic
behavior \eqref{leading} and the third line reflects that only the
first term in \eqref{f-rec} is relevant when $x_d\gg x_i$ for all
$i<d$. Our last step is to multiply and divide the third line by $x_d$
and then invoke the scaling law \eqref{tot} for the average number of
$k-1$-minima in $d-1$ dimensions.  In essence, we utilize the fact
that when one of the coordinates is very large, the partial minima
criterion involves one less constraint in one less dimension
\cite{one}. The power-law decay of the distribution \eqref{largest}
shows that there is a substantial likelihood that $x_d$ is relatively
large.

The distribution $Q_2(x_{d-1})$ of the second largest coordinate
$x_{d-1}$ is obtained using the bivariate distribution $\tilde
Q(x_{d-1},x_d)$,
\begin{eqnarray}
\label{bivariate}
\tilde Q(x_{d-1},x_d)\!
&\!=\!&\!\int dx_1\cdots \int dx_{d-2}\,P_{d,k}(x_1,x_2,\cdots,x_d)\nonumber\\
&\!\sim\!&\!\int dx_1\cdots \int dx_{d-2}\, Ne^{-NF_{d,k}({\bf x})}\nonumber\\
&\!\sim\!&\!\int dx_1\cdots\int dx_{d-2}\, Ne^{-Nx_{d-1}x_dF_{d-1,k-1}({\bf x})}\nonumber\\
&\!\sim\!&\!N^{-\frac{d-k}{k-2}}(x_{d-1}x_d)^{-\frac{d-k}{k-2}-1}.
\end{eqnarray}
The distribution $Q_2(x_{d-1})$ equals the integral of the bivariate
distribution with respect to the largest coordinate,
$Q_2(x_{d-1})=\int_{x_{d-1}}^1 dx_d \,\tilde Q(x_{d-1},x_d)$. This
integral is dominated by the divergence at the lower limit of
integration, and consequently 
\begin{equation}
Q_2(x_{d-1})\sim N^{-\frac{d-k}{k-2}}(x_{d-1})^{-2\frac{d-k}{k-2}-1}.
\end{equation}
The power-law tail is now steeper.

A similar calculation applies to the distributions of the $k-1$
largest elements.  In general, the distribution $Q_j(y_j)$ of the
$j$th largest element, $y_j$, with the definition \hbox{$y_j\equiv
x_{d+1-j}$}, decays as a power-law,
\begin{equation}
Q_j(y_j)\sim N^{-\frac{d-k}{k-j}}(y_j)^{-\alpha_j-1}
\end{equation}
for $1\leq j\leq k-1$. The decay exponent increases monotonically with
the index $j$,
\begin{equation}
\label{alphaj}
\alpha_j=j\frac{d-k}{k-j}.
\end{equation}
We can verify the decay law \eqref{average} using \hbox{$A\sim
\int_{N^{-1/k}}^1 dy_j\, Q_j(y_j)$} where the lower limit of
integration is set by the typical size scale
\eqref{scale}. Interestingly, there are $k-1$ distinct scaling
behaviors for the $k-1$ largest elements. Each of these extremal
coordinates is distributed according to a power-law distribution that
is characterized by a distinct exponent.

This multiscaling behavior affects the behavior of the moments
$\langle y_j^m\rangle$ defined as follows, $\langle y_j^m\rangle
=I_m/I_0$, where $I_m=\int_{N^{-1/k}}^1 dy_j \,y_j^m \,Q_j(y_j)$. The
integral $I_m$ is dominated by the divergence at the lower cutoff when
the order is small, $m\leq \alpha_j$, but otherwise, the integral
$I_m$ is finite. Consequently, the moments have the following scaling
dependences on $N$
\begin{eqnarray}
\langle y_j^m\rangle \sim
\begin{cases}
N^{-m/k}& m<\alpha_j,\\
\frac{1}{k}N^{-\frac{j(d-k)}{k(k-j)}}\ln N& m=\alpha_j,\\
N^{-\frac{j(d-k)}{k(k-j)}}& m>\alpha_j.\\
\end{cases}
\end{eqnarray}
Low order moments exhibit ordinary scaling behavior as they are
characterized by the typical size scale \eqref{scale} that underlies
the multivariate distribution function \eqref{scaling-function}. As
usual, there is a logarithmic correction at the crossover.  High-order
moments plateau at a fixed value that is independent of the index $m$,
an indication that there is a significant probability that the extreme
elements are of order one.  Interestingly, the average size of the
different coordinates may follow different scaling laws. For example,
there are two scaling laws, $\langle y_1\rangle \sim N^{-1/6}$ and
$\langle y_2\rangle \sim N^{-1/3}$ when $d=4$ and $k=3$. Of course,
the sum $\sum_{i=1}^d x_i$ has the same extremal statistics as does
$x_d$.

The crossover moment or equivalently the exponent $\alpha_j$ diverges
as $k\to j$. Therefore, the smallest $d+1-k$ coordinates exhibit the
ordinary scaling behavior
\begin{equation}
\langle y_j^m\rangle \sim N^{-m/k}
\end{equation}
for $k\leq j\leq d$ and all moments of the respective distribution
functions must be finite. In these cases, the distribution functions
$Q_j$ have tails that are as sharp as or sharper than an
exponential. In the aforementioned case $d=4$ and $k=3$, the third and
the fourth largest coordinates exhibit the ordinary scaling, $\langle
y_3\rangle \sim \langle y_4\rangle \sim N^{-1/3}$.

\noindent{\bf Efficient Sets.}  The set of points that are not
dominated on all coordinates by any other point are partial minima
when $k=d$ (figure 1). We refer to this set as the ``efficient set''.
The efficient set, also termed the efficient frontier or Pareto
equilibria, plays a central role in multi-objective optimization and
has been studied in economics, computer science, operations research,
and game theory because every point in the set is a candidate solution
to the multi-objective optimization problem, depending on the relative
weights of the various costs \cite{osw,fw}.

\begin{figure}[t]
\includegraphics*[width=0.25\textwidth]{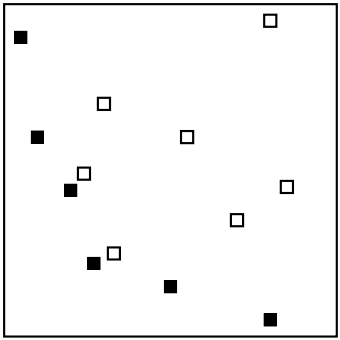}
\caption{Illustration of the efficient set in two-dimensions.  Filled
squares are on the efficient set and unfilled squares are not. Only
four of the filled squares are on the convex hull.}
\end{figure}

In the special case $k=d$, the expected size of the efficient set,
$E_d(N)\equiv A_{d,d}(N)$, obeys the recursion
\begin{equation}
\label{e-rec}
E_d(N)=E_d(N-1)+\frac{1}{N}\,E_{d-1}(N).
\end{equation}
The point with the largest $x_d$ coordinate certainly does not
dominate any other point. Furthermore, this point is on the efficient
set if and only if the rest of its $d-1$ coordinates are not dominated
by any other point.  This event occurs with probability
$\frac{1}{N}E_{d-1}(N)$ and hence, the second term in the
recursion. We note that the recursion \eqref{e-rec} can also be
obtained by performing the integration over $x_d$ in $E_d(N)=N\int
d{\bf x}\,[1-x_1x_2\cdots x_d]^{N-1}$.  This integration is
analytically feasible only if $k=1$ or $k=d$.

The recursion relation \eqref{e-rec} is subject to the boundary
condition $E_1(N)=1$.  In two dimensions,
\begin{equation}
E_2(N)=1+\frac{1}{2}+\frac{1}{3}+\cdots+\frac{1}{N},
\end{equation}
or alternatively, $E_2(N)=H(N)$, where $H(N)=\sum_{n=1}^N \frac{1}{n}$
is the harmonic number.  The average size of the efficient set grows
logarithmically, $E_2(N)=\ln N+\gamma+\cdots$ where $\gamma=0.57721$
is Euler's constant. In three dimensions, we have
\hbox{$E_3(N)=\sum_{n=1}^N\frac{1}{n}H(n)$}, and asymptotically,
\hbox{$E_3(N)\simeq \frac{1}{2}(\ln N)^2$}. The large-$N$ behavior is
obtained in general by converting the difference equation
\eqref{e-rec} into a differential equation
\hbox{$dE_d/dN=E_{d-1}/N$}. The expected size of the efficient set
grows logarithmically,
\begin{equation}
\label{log}
E_d(N)\simeq \frac{1}{(d-1)!}(\ln N)^{d-1}.
\end{equation}

This logarithmic growth reflects that the integral of the scaling
function, $\int d{\bf z}\,\Phi_{d,d}({\bf z})$, is divergent at the
upper limit.  A straightforward generalization of the calculation
above shows that the distribution of the extremal coordinates has a
logarithmic correction,
\begin{equation}
Q_j(y_j)\sim (\ln N)^{d+1-j}(y_j)^{-1}|\ln y_j|^{j-1},
\end{equation}
for $1\leq j\leq d-1$.  We can verify that the average number of
points is consistent with the exact behavior $\int_{N^{-1/d}} dy\,
Q_j(y_j) \sim (\ln N)^{d-1}$ as in \eqref{log}.  The crossover moment
vanishes and the moments decay logarithmically,
\begin{equation}
\langle y_j^m\rangle \sim (\ln N)^{-j},
\end{equation}
where $m>0$ and $1\leq j\leq d-1$.

\noindent{\bf Two-dimensions.}  In two-dimensions, the full
probability distribution function $p_n(N)$ that the efficient set
includes $n$ points, where $1\leq n\leq N$, satisfies the recursion
\cite{extremes}
\begin{equation}
\label{p-rec}
p_n(N)=\left(1-N^{-1}\right)p_n(N-1)+N^{-1}p_{n-1}(N-1)
\end{equation}
and is subject to the boundary condition $P_n(0)=\delta_{n,0}$. On the
square, there are two coordinates: $x_1$ and $x_2$. Following the
reasoning behind \eqref{e-rec}, the point with the largest $x_2$
coordinate is on the efficient set if and only if its $x_1$ coordinate
is minimal, an event that occurs with probability $N^{-1}$.

Recursion equations for the average \hbox{$E(N)=\langle n\rangle$} and
the variance \hbox{$V(N)=\langle n^2\rangle -\langle n\rangle^2$} with
\hbox{$\langle f(n)\rangle\equiv \sum_{n=1}^N f(n)P_n$} are obtained
by summing \eqref{p-rec}. The average satisfies $E(N)=E(N-1)+N^{-1}$
in accord with \eqref{e-rec} and the variance satisfies
\hbox{$V(N)=V(N-1)+N^{-1}-N^{-2}$}. Thus, the variance equals the
difference between the first and the second harmonic numbers
\begin{equation}
V(N)=H(N)-H^{(2)}(N)
\end{equation}
where $H^{(2)}(N)=\sum_{n=1}^N n^{-2}$. The variance and the average
have identical leading asymptotic behaviors, \hbox{$V(N)=\ln
N+(\gamma-\frac{1}{6}\pi^2)+\cdots$}.

With the transformation $p_n(N)=\frac{1}{N!}\tilde p_n(N)$, the
auxiliary function $\tilde p_n(N)$ satisfies the recursion
\begin{equation}
\tilde p_n(N)=(N-1)\tilde p_n(N-1)+\tilde p_n(N-1)
\end{equation}
with $\tilde p_n(0)=\delta_{n,0}$. This recursion defines the Stirling
numbers ${N\brack n}$ \cite{gkp} so $\tilde p_n(N)={N\brack
n}$. Therefore, the full probability distribution is expressed in
closed form,
\begin{equation}
\label{dis-sol}
p_n(N)=\frac{1}{N!}\,{N\brack n}
\end{equation}
for $0\leq n\leq N$.

The general asymptotic behavior, derived in \cite{bk},
\begin{equation}
\label{modified}
p_n(N)\simeq \frac{1}{N}
\frac{1}{\Gamma(n/\ln N)}\,
\frac{(\ln N)^n}{n!}
\end{equation}
applies in the limit $n\to\infty$ $N\to\infty$ with the ratio $n/\ln
N$ finite. For small $n\ll \ln N$, the distribution is Poissonian,
$P_n(N)=N^{-1}(\ln N)^{n-1}/(n-1)!$ and for large $n$, the
distribution approaches a Gaussian centered at the average $E(N)\simeq
\ln N$ with the variance $V(N)\simeq \ln N$,
\begin{equation}
\label{gaussian}
p_n(N)\to \frac{1}{\sqrt{2\pi \ln N}}\,
\exp\left[-\frac{(n-\ln N)^2}{2\ln N}\right].
\end{equation}
We note that the convex hull, a subset of the efficient set (see
figure 1), is characterized by similar statistical properties
including a limiting Gaussian distribution and logarithmic growths,
albeit with different prefactors, of the average and the variance
\cite{ars,be,vv}.

\noindent{\bf Multi-Objective Shortest Path.} The multi-objective
shortest path on a graph is defined as follows.  Consider a graph,
possibly with multiple edges connecting pairs of nodes, with $d$
different costs on each edge.  Fix the source and the destination
nodes, and then consider all possible paths from source to
destination, assigning a vector of $d$ total costs to each path.  The
multi-objective shortest path problem requires identification of the
efficient set of paths.  Generally, finding the efficient set is an
NP-hard problem, although efficient approximation schemes exist
\cite{approx1,approx2}.  The computation time of the approximation
scheme depends crucially on the size of the efficient set.

Suppose the weights are assigned independently at random to the
edges. We can consider two limiting topologies.  First, for a graph of
two nodes connected by $N$ edges, the efficient set grows only
polylogarithmically in the number of edges following the calculation
above. Second, for a one-dimensional chain of nodes where each pair of
neighboring nodes is connected by a pair of edges, the weights of the
paths become correlated \cite{approx1}. We have conducted numerical
studies, and found that the size of the efficient set is highly
sensitive to the distribution of weights on the edges. Assuming each
edge has two weights, $(w_1,w_2)$, both chosen from some continuous
distribution, the convex hull grows linearly in the length of the
chain. Interestingly, we observed various behaviors for the size of
the efficient set, ranging from linear in the length of the chain, to
power law behavior with various exponents greater than unity, up to
stretched exponential behavior.

Finally, consider an Erd\"{o}s-Renyi random graph of $M$ nodes
\cite{bb,jlr}.  Two randomly chosen nodes will typically have a
shortest path distance between them of order $\log M$ using a metric
which simply counts the number of edges traversed.  While it is
possible for a path on the efficient set to be longer than this,
because the weights are positive, we expect that paths on the
efficient set will be at most of order $\sqrt{\log M}$ longer.  The
total number of paths of at most that length grows exponentially in
$\sqrt{\log M}$, and such paths will tend to overlap only near the
source and destination nodes.  Thus, to a good approximation we expect
that the weights of the paths will be uncorrelated, enabling us to use
the results above.  Then, the number of paths on the efficient set
will only be of order $(\log M)^{d/2}$.  In general, then, when the
paths have little overlap as here, the number of paths on the
efficient set is much smaller than in cases like one-dimension where
the paths greatly overlap.

\noindent{\bf Conclusions.} We studied statistical properties of
partial minima in a set of uncorrelated points in general
dimensions. These partial minima are defined by a parameter $k$: a
point is a partial minimum if it dominates all other points on at
least $d-k$ coordinates. As this condition becomes more stringent,
partial minima improve in quality but are less probable.  Remarkably,
there is a series of distinct power-law distributions that characterize
the largest coordinates with a consequent multiscaling distribution of
the moments, while the rest of the coordinates obey ordinary
scaling. In the extreme case $k=d$, the number of partial minima grows
logarithmically with the total number of points.

Our results hold as long as the set of points are not correlated, that
is, as long as they are drawn from independent distributions.  These
distributions need not be identical. If the $i$th coordinate is drawn
from the distribution $f_i(x_i)$, the transformations $x_i\to
\int_0^{x_i} dy_i f_i(y_i)$ and $dx_i\to f_i(x_i)dx_i$, maps to a
uniform distribution in the unit hypercube. Correlations present an
interesting challenge and we anticipate serious modifications to the
scaling laws above.  For instance, it is simple to show that the size
of the efficient set grows as a power of the number of points, $\sim
N^{1/2}$, rather than a logarithm, when the points are uniformly
distributed inside the unit circle. Incidentally, this growth is much
faster than the $N^{1/3}$ for the corresponding number of points in
the convex hull \cite{ars}.

Another interesting issue is the crossover from the algebraic decay
\eqref{average} to the logarithmic growth \eqref{log}.  The average
number of partial minima decreases monotonically with $N$ when $k$ is
small, but is a non-monotonic function of $N$ when $k$ is large. For
example, when $d=4$ and $k=3$, the average $A_{d,k}$ peaks at
$N=16$. It will be interesting to elucidate how the height and the
location of this peak scales with $N$.

\noindent{\bf Acknowledgments.} We thank Gunes Ercal-Ozkaya for
suggesting random graphs and Paul Krapivsky for useful discussions on
convex hulls.  We acknowledge financial support from DOE grant
DE-AC52-06NA25396.


\begin{thebibliography}{99}

\bibitem{rs}
      R.~E.~Steuer, {\it Multiple Criteria Optimization:
      Theory, Computations, and Application} (Wiley, New York, 1986).

\bibitem{snt}
       Y.~Sawaragi, H.~Nakayama, and T.~Tanino,
       {\it Theory of Multiobjective Optimization},
       (Academic Press, Orlando, 1985).

\bibitem{ft}
       D.~Fudenberg and J.~Tirole,
       {\it Game Theory},
       (MIT Press, Cambridge, 1983)

\bibitem{or}
        M.~J.~Osborne and A.~Rubenstein,
    {\it A Course in Game Theory},
    (MIT Press, Cambridge, 1994).

\bibitem{sg} M.~M\'ezard, G.~Parisi, and M.~Virasoro, {\it Spin Glass
    Theory and Beyond} (World Scientific, Singapore, 1987).

\bibitem{dl}
    B.~Derrida and J.~L.~Lebowitz, 
    Phys.\ Rev.\ Lett. {\bf 80}, 209 (1998).

\bibitem{bkm}
   E.~Ben-Naim, P.~L.~Krapivsky, and S.~N.~Majumdar,
   Phys.\ Rev.\ E {\bf 64}, 035101 (2001).

\bibitem{ktnr} G.~Korniss, Z.~Toroczkai, M.~A.~Novotny, and P.~A.~Rikvold,
      Phys. Rev. Lett. {\bf 84}, 1351 (2000).

\bibitem {kr}
   P.~L.~Krapivsky and S.~Redner,
   Phys.\ Rev.\ Lett.\ {\bf 89}, 258703 (2002).

\bibitem{rec} For small $N$, these integrals can be calculated
     manually or through recursion formulas. For example,
     \hbox{$A_{d,k}=\frac{1}{2}(A_{d-1,k-1}+A_{d-1,k-1})$} when $N=2$.

\bibitem{ejg}
   E.~J.~Gumbel, {\em Statistics of Extremes} (Columbia University
   Press, New York, 1958).

\bibitem{jg}
   J.~Galambos, {\em The Asymptotic Theory of Extreme Order Statistics}
   (R.E. Krieger Publishing Co., Malabar, 1987).

\bibitem{res}
   R.~E.~Ellis, {\em Entropy, Large Deviations, and Statistical Mechanics}
   (Springer-Verlag, New York, 1985).

\bibitem{one}
       Formally, when $x_d=1$ then $G_{d,k}({\bf x})=G_{d-1,k-1}({\bf x})$.

\bibitem{osw}
      T.~Ottmann, E.~Soisalon-Soininen, and D.~Wood,
      Info. Sci. {\bf 33}, 157 (1984).

\bibitem{fw}
      E.~Fink and D.~Wood,
      Jour. Geometry {\bf 62}, 99 (1998).

\bibitem{extremes} Generally, $p_1(N)=A_{d,1}=N^{-(d-1)}$.

\bibitem{gkp}
     R.~L.~Graham, D.~E.~Knuth, and O.~Patashnik,
     {\it Concrete Mathematics: A Foundation for Computer Science}
     (Reading, Mass.: Addison-Wesley, 1989).

\bibitem{bk}
     E.~Ben-Naim and P.~L.~Krapivsky,
     J. Phys. A {\bf 37}, 5949 (2004).

\bibitem{ars}
     A.~R\'enyi and R.~Sulanke,
     Z. Wahrsch. Verw. Geb.  {\bf 2},  75--84 (1963).

\bibitem{be}
     B.~Efron,
     Biometrika, {\bf 52}, 331 (1965).

\bibitem{vv}
     V.~Vu,
     Adv.\ Math. {\bf 207}, 221--243 (2006).

\bibitem{approx1} 
     A. Warburton, 
     Oper. Res. {\bf 35}, 70 (1987).

\bibitem{approx2} 
     G. Tsaggouris and C. Zaroliagis,
     Lect.  Notes in Comp. Sci. {\bf 4288}, 389 (2006).

\bibitem{bb}
   B.~Bollob\'as,
   {\em Random Graphs} (Academic Press, London, 1985).

\bibitem{jlr}
   S.~Janson, T.~\L uczak, and A.~Rucinski,
   {\em Random Graphs} (John Wiley \& Sons, New York, 2000).


\end{thebibliography}
\end{document}